\documentclass[12pt]{article}
\usepackage{pazha}
\usepackage{graphicx}
\usepackage{lscape}
\usepackage{amssymb}
\usepackage{amsmath}

\tightenlines
\def\ÐÍ{$\pm$}

\def\arcmin{$^{\prime\,}$}

\begin{document}

{\it Will be published in ``Astronomy Letters'', 2009, v.35, N4, pp.
235-240}

\bigskip

\title{\bf Can the Known Millisecond Pulsars Help in the Detection
of Intermediate-Mass Black Holes at the Centers of Globular
Clusters?}

\author{\bf T.I.Larchenkova\affilmark{1,*} and A.A.Lutovinov\affilmark{2,**}}

\affil{ {\it Astro Space Center, Lebedev Physical Institute, Russian
Academy of Sciences, Profsoyuznaya str., 84/32, Moscow, 117997
Russia}$^1$\\
{\it Space Research Institute, Russian Academy of
Sciences, Profsoyuznaya str., 84/32, Moscow, 117997 Russia}$^2$\\ }
\vspace{2mm}

\received{14 May 2008}

\sloppypar
\vspace{2mm}
\noindent

We consider the possibility of detecting intermediate-mass
($10^3-10^4 M_{\odot}$) black holes, whose existence at the centers
of globular clusters is expected from optical and infrared
observations, using precise pulse arrival timing for the millisecond
pulsars in globular clusters known to date. For some of these
pulsars closest to the cluster centers, we have calculated the
expected delay times of pulses as they pass in the gravitational
field of the central black hole. The detection of such a time delay
by currently available instruments for the known pulsars is shown to
be impossible at a black hole mass of $10^3 M_{\odot}$ and very
problematic at a black hole mass of $10^4 M_{\odot}$. In addition,
the signal delay will have a negligible effect on the pulsar periods
and their first derivatives compared to the current accuracy of
their measurements.

\vspace{10mm} Key words: globular clusters, black holes, pulsars,
Shapiro effect

\vfill

\noindent
$^*$ e-mail: tanya@lukash.asc.rssi.ru\\
$^{**}$ e-mail: aal@iki.rssi.ru

\clearpage

\section*{INTRODUCTION}

\vskip -5pt

The question of whether black holes with masses $10^3–10^4
M_{\odot}$ are present at the centers of globular clusters arose in
the 1970s as new X-ray observational data became available (see,
e.g., Clarket al. 1975). In subsequent theoretical works, the effect
of a massive black hole in a dense stellar system on its dynamical
and astrophysical properties were discussed and numerical solutions
were obtained for some stationary cases (see, e.g., Frank and Rees
1976; Bisnovatyi-Kogan et al. 1980). In addition, for some of the
clusters, for example, for the globular cluster M15, the detection
of a black hole at its center was reported; subsequently,this report
was disproved (for the history of this question, see, e.g., McNamara
et al. 2003).

In recent years, evidence for the existence of intermediate-mass
blackholes (IMBHs), at least in two globular clusters, $\omega$Cen
(NGC 5139) and G1 (Mayall II) in the galaxy M31 (Noyola et al. 2006;
Gebhardt et al. 2005), has appeared in connection with an
improvement in observational optical and infrared instruments. In
particular, the black hole in the latter cluster was detected and
its mass was determined on the basis of Hubble Space Telescope
(HST)photometry and HIRES (Keck telescope) spectroscopy (Gebhardt et
al. 2005). The observational data were used to construct the
dependence of the mass to-luminosity ($M/L$) ratio on the distance
to the cluster center. This dependence shows that the $M/L$ ratio
increases significantly at small cluster radii. This behavior of
$M/L$ is difficult to explain by the presence of low-luminosity
stellar objects near the cluster center, such as white dwarfs and
neutron stars. However, it can be well described in terms of
dynamical models for globular clusters with massive
($\sim2\times10^4 M_{\odot}$) central objects (Gebhardt et al.
2005). Apart from the clusters mentioned above, there is
circumstantial evidence for the existence of a central blackh ole in
the globular clusters 47 Tuc, NGC 6752, and M15 (McLaughlin et al.
2006; van den Bosch et al. 2006). XMM-Newton X-ray observations
revealed a black hole with a mass of $\sim10^3 M_{\odot}$ in a
globular cluster of the elliptical galaxy NGC 4472, which, in turn,
belongs to the Virgo cluster of galaxies (Maccarone et al. 2007). It
is important to note that the IMBHs at the centers of globular
clusters in our Galaxy may no manifest themselves in X rays. As was
shown by Frank and Rees (1976), the disruption rate of stars by
black hole with a mass of $10^3-10^4 M_{\odot}$ is low. Therefore,
all of the possible methods for detecting these objects in all of
the wavelength ranges accessible to observers should be used,
because the problem of the existence of IMBHs is important in
understanding not only the structure and evolution of globular star
clusters but also the evolution of galaxies with central black holes
and the formation of such objects.

Note that the above-mentioned circumstantial evidence for the
existence of IMBHs has been obtained for massive globular clusters
with compact dense cores. The largest number of radio pulsars and
Xray binaries have been discovered precisely in these globular
clusters. In particular, the globular clusters 47 Tuc and M15 that
are suspected to host a central blac khole contain 22 and 10 radio
pulsars, respectively
(http://www.naic.edu/$\sim$pfreire/GCpsr.html).

The radio pulsars have a remarkable property, a high stability of
their pulsed radiation, which can be used to independently confirm
the existence of a central black hole in a globular cluster. If a
massive compact object is located near the propagation ray of the
signal from the pulsar to the observer, then the flux from the
pulsar will be magnified (or, in some cases, attenuated) due to
gravitational lensing. However, apart from the flux magnification, a
signal time delay will also be observed in this case (Krauss and
Small 1991). Both these effects can be used to detect the black
holes located at the centers of globular clusters. However, the
gravitational lensing of Galactic sources (stars and pulsars) in the
gravitational field of such black holes is difficult to observe,
since the Einstein-Chwolson radius is fairly small ($\sim10$ AU),
while for the flux from the source to be amplified, the line of
sight must pass within this small radius (see, e.g., Zakharov 1997).

The angular distances of the currently known pulsars in globular
clusters from their centers are typically a few hundredths of an
arcminute, which is much larger than the Einstein-Chwolson angular
radius.As will be shown below, the relativistic time delay of the
pulsar electromagnetic radiation, called the Shapiro effect, may
turn out to be significant in this case. An expression for this
effect was derived for an electromagnetic signal propagating in a
static, spherically symmetric gravitational field of a point mass by
Shapiro (1964). Kopeikin and Schafer (1999) generalized the
expression to the case of light propagation in a variable field of
an arbitrary moving body.

Wex et al. (1996) showed that the timing of millisecond pulsars can
be used in principle to detect and identify massive objects in the
Galactic Center region. In this paper, based on the calculations of
the above paper, we investigate the possibility of detecting IMBHs
at the centers of globular clusters using long-term observations of
currently known pulsars located at minimum angular distances from
the cluster center. Note that the Shapiro effect was previously
suggested to be used as a possible cause of the pulsar glitches
(Sazhin 1986), to search for dark-matter objects in the Galaxy
(Larchenkova and Doroshenko 1995), to test the general theory of
relativity in binary pulsars (Doroshenko and Kopeikin 1995), to
probe the structure of the Galaxy, and to search for compact
objects(Larchenkova and Lutovinov 2007, Siegel 2008).\\


\section*{FORMULAS AND DEFINITIONS}

Following the notation used previously (Larchenkova and Lutovinov
2007), Fig. 1 presents a classical model of gravitational lensing
for a point lens (a black hole) with mass $M_{BH}$. The position of
the pulsar (PSR) in the sky relative to the observer (O) is
specified by the angle $\theta_s$, while the positions of its images
are specified by the angles $\theta_{+}$ and $\theta_{-}$, where the
'$+$' and '$-$' signs correspond to the first ($+$) and second ($-$)
images, respectively, and $d$ is the impact parameter of the
undeflected light ray. In this case, the Einstein-Chwolson radius is
defined by the formula

\begin{equation}
R_E=(4GM_{BH}D_{ds}D_d/c^2 D_s)^{1/2},
\end{equation}

\noindent where $c$ is the speed of light in a vacuum, $G$ is the
gravitational constant, $D_{ds}$ and $D_d$ are the distances from
the pulsar to the black hole with mass $M_{BH}$ and from the black
hole the observer, respectively, $D_s$ is the distance from the
pulsar to the observer, and $D_s=D_{ds}+D_d$ (the formula for the
radius $R_E$ was first derived by Einstein (1965, vol. 2) for
$D_{ds}\gg D_d$; Eq. (1) in the commonly used form can be found,
e.g., in Vietri and Ostriker (1983)). Note that when the pulsar and
the black hole are located in the same globular cluster, $R_E
\propto D_{ds}^{1/2}$ with a high accuracy, since the typical size
of globular clusters (several tens of parsecs) is much smaller than
the distance from these clusters to the observer (several kpc). In
general form, the dependence $R_E(D_{ds})$ and its typical values
are presented in Fig. 2 for two black hole masses, $M_{BH} = 10^3$
and $10^4 M_{\odot}$.

The angular distance between the two images, ($+$) and ($-$), is
defined as (Refsdal 1964; Turner et al. 1984)

\begin{equation}
 \Delta\theta = \frac{R_E}{D_d}\sqrt{f^2+4},
\end{equation}
\noindent where $f$ is the dimensionless impact parameter, $f =
d/R_E$. For example, the angular distance between the images for
pulsars in the globular cluster M15 ($D_d = 10.2$ kpc and we set
$D_{ds}$ equal to $\sim3$ pc, which corresponds to the typical radii
$r_h$ within which half of the cluster mass is concentrated) is
$\Delta\theta=1.533\times10^{-5}\sqrt{M_{BH,M_{\odot}}}\sqrt{f^2+4}$
arcmin, where $M_{BH,M_{\odot}}$ is the black hole mass in solar
masses.

Apart from the appearance of two images in the plane of the
gravitating body, flux magnification and a signal time delay must
also be observed in the classical model of gravitational lensing.
The former is specified by the formula (Refsdal 1964; Wex et al.
1996)

\begin{equation}\label{eq_mu}
 \mu_{+,-}=\frac{1}{4}\left[\frac{f}{(f^2+4)^{1/2}}+
 \frac{(f^2+4)^{1/2}}{f} \pm 2\right]\ .
\end{equation}
\noindent

It follows from Eq. (3) that the contribution from the second ($-$)
image to the total brightness is small ($\leq3$\%) even at $f\geq2$
(below, we will show that the value of this quantity is much higher
for the known pulsars) and this image is too faint to be observable.
All of the subsequent reasoning refers only to the first ($+$)
image.

For a spherically symmetric Schwarzschild lens, the signal delay can
be expressed as (see, e.g., Krauss and Small, 1991)

\begin{equation}\label{eq_tau}
\tau_{+,-} = \frac{2GM}{c^3} \left[ \frac{4}{(\sqrt{f^2+4}\pm f)^2}
- \ln(\sqrt{f^2+4} \pm f)^2 \right]+const
\end{equation}

\noindent where the first and second components reflect the
geometric and relativistic time delays, respectively. The delay
depends on the impact parameter, which, in turn, varies with time
due to the relative motion of the pulsar and the black hole:

$$f=\frac{d}{R_E}=\frac{d_m}{R_E}\sqrt{1+\left(\frac{v_{\perp}}{d_m}\right)^2(t-T_0)^2}.$$

Here, $v_{\perp}$ is the pulsar velocity relative to the black hole
projected onto the plane of the sky, $d_m$ is the minimum impact
parameter, $t$ is the current observation time, and $T_0$ is the
time of the closest approach. Thus, the change in delay $\Delta
\tau=\tau(t)-\tau(t_0)$, where $t_0$ is the time at which the
observations begin, is a measurable quantity. In Fig. 3, $\Delta
\tau$ is plotted against the observation time for the minimum impact
parameter $d_m=10^{4}$ AU (below, we will show that the observed
pulsars in globular clusters are located at similar or larger
distances from their centers), the velocity $v_{\perp}=30$ km
s$^{-1}$ (a typical velocity of globular-cluster stars), $t_0-T_0 =
5$ yr, and two black hole masses, $M_{BH}=10^3$ and $10^4
M_{\odot}$. We see that the maximum signal delay in this case is
small, $\sim100$ ns and $\sim1~\mu$s, respectively.

\section*{KNOWN PULSARS IN GLOBULAR CLUSTERS}

Let us now use the above reasoning and formulas to assess the
observability of the delay of a signal as it passes near an IMBH for
several known pulsars detected in globular clusters. For this
purpose, we took the pulsars from Freire's catalog
(http://www.naic.edu/$\sim$pfreire/GCpsr.html) closest to the
centers of the globular clusters suspected to host black holes.
Assuming that the pulsar is located behind the cluster center at a
distance of 3 pc, its transverse velocity is 30 km s$^{-1}$, and the
observation time is 5 yr, we calculated the maximum relative delay
of its signal as it passed near a black hole with masses
$M_{BH}=10^3$ and $10^4 M_{\odot}$ for each pulsar ($\Delta\tau_3$
and $\Delta\tau_4$, respectively). The pulsar parameters (the offset
-- the angular distance between the pulsar and the cluster center)
and the results of our calculations are given in the table.

We see from the table that at the currently achievable accuracy of
determining the pulse arrival time (PAT) ($\sim50$ ns for bright
pulsars), the observation of the signal time delay for the known
millisecond pulsars as their signals pass near the black holes at
the centers of the corresponding globular clusters is not possible
at a black hole mass of $10^3 M_{\odot}$ and is very problematic at
a black hole mass of $10^4 M_{\odot}$ even for pulsars close to the
cluster center (e.g., B2127+11D or J0024-7204O). A two-fold increase
in the duration of observations, to 10 yr, leads to a significant
increase in the observed relative signal delay, which can already be
detected by currently available instruments. It should be noted that
the low-frequency noise due to the motion of globular-cluster stars
will have a significant effect on the detectability of single pulse
delay events related to the passage of the pulsar signal near a
massive black hole (for more detail, see Kopeikin 1999; Larchenkova
and Kopeikin 2006).

\subsection*{Effects of the Pulsar Signal Delay on the Observed
Periods and Their Derivative}

Let us consider how the time delay of the pulsar signal as it passes
near a black hole will affect the observed pulsar period and its
first derivative and whether a conclusion about the presence of an
IMBH at the cluster center can be drawn from their measurements for
the pulsars known to date. For this purpose, we will use reasoning
similar to that in Wex et al. (1996). The observed pulsar period $P$
at time $t_1$ is related to the intrinsic pulsar period $P_i$ as

\begin{equation}\label{eq_per}
 P\simeq P_i+\frac{d\tau}{dt}(t_1).
\end{equation}
\noindent

As was noted above, only one ($+$) image is observed in the case of
''weak lensing'' ($f\gg1$) and the geometric delay is negligible
(see Eq. (4)). In this case, the time delay is determined only by
the Shapiro effect and can be written as (Larchenkova and Doroshenko
1995)

\begin{equation}\label{eq_tau_shap}
\tau_{+} = - \frac{2GM}{c^3}
             \ln\left(1+\left(\frac{v_{\perp}}{d_m}\right)^2 \left(t-T_0\right)^2\right) .
\end{equation}

Substituting Eq. (6) into (5) and searching for a maximum of the
derived function, we find (Wex et al. 1996)

\begin{equation}\label{eq_pmax}
{\rm max}\left| \frac{P}{P_i}-1 \right| \simeq \frac{2GM}{c^3}
\frac{v_{\perp}}{d_m}.
\end{equation}

For the pulsar B2127+11D from the table, which is closest to the
center of the globular cluster M15, the maximum changes in pulsation
period are

$${\rm max}\left| \frac{P}{P_i}-1 \right| \simeq 1.7\times10^{-13} ~~ {\text and} ~~ 1.7\times10^{-12}$$

\noindent for a black hole mass of $10^3$ and $10^4 M_{\odot}$,
respectively. The typical accuracies of measuring the periods of
millisecond pulsars in globular
clusters\footnote{http://www.atnf.csiro.au/research/pulsar/psrcat}
are several orders of magnitude lower than the above estimates of
the maximum change in pulsar period, which, in addition, can be
achieved over a disproportionately long time of source observations
($\sim10^3$ yr).

Let us now consider how the signal delay affects the first
derivative of the pulsation period. Let again $\dot P$ be the
measurable rate of change in pulsar period and $\dot P_i$ be the
intrinsic change in period. Using the same approach as that for the
pulsation period, we find in general form that (Wex et al. 1996)

\begin{equation}\label{eq_pdotmax}
{\rm max}\left| \frac{\dot P - \dot P_i}{P_i} \right| \simeq
\frac{4GM}{c^3} \left(\frac{v_{\perp}}{d_m}\right)^2,
\end{equation}
\noindent and for the pulsar B2127+11D

$${\rm max}\left| \frac{\dot P - \dot P_i}{P_i} \right| \simeq 5.8\times10^{-24} {\text c}^{-1} ~~ {\text and} ~~ 5.8\times10^{-23} {\text c}^{-1}$$
\noindent at a black hole mass $10^3$ and $10^4 M_{\odot}$,
respectively. These values are again several orders of magnitude
lower than the typical accuracies of measuring the first derivative
of the millisecond pulsar period. Thus, the effect of an IMBH on the
periods and their derivatives for the known millisecond pulsars in
globular clusters is too weak to be detected by currently available
instruments.

\section*{A PULSAR IN A GALAXY BEHIND A GLOBULAR CLUSTER}

In conclusion, let us consider the hypothetical case where a
millisecond pulsar is located behind a globular cluster at large
(several kpc) and small angular distances from the cluster center.
Clearly, the probability of such a case is low. Nevertheless, there
are observational examples of such a mutual arrangement of objects
even now (in particular, the pulsar J1748-2446B, later renamed as
PSR J1744-2444, in the globular cluster Ter 5,
http://www.naic.edu/$\sim$pfreire/GCpsr.html). For the subsequent
estimates, we will assume this angular distance to be the same as
that for the pulsar B2127+11D ($0.019$\arcmin) and the distance
between the globular cluster (M15) and the pulsar to be 3 kpc (i.e.,
the pulsar is located somewhere in the halo, at the edge of the
Galaxy). We see from Fig. 2 that the Einstein-Chwolson radius for
such distances increases to several hundred AU. For our estimate, we
will take the velocity of the halo and Galactic objects to be
$\sim200$ km s$^{-1}$. The maximum relative signal delay recorded
over 5 yr is then $\sim2~\mu$s for a black hole mass of $10^3
M_{\odot}$ and an order of magnitude larger for a mass of $10^4
M_{\odot}$. Note that the significant increase in signal delay
depends weakly on the distance between the globular cluster and the
pulsar but is related to considerably higher velocities of the
Galactic objects than those of the globular-cluster objects. In this
case, the effect of the delay on the pulsation period will still be
weak due to the large minimum impact parameter $d_m$ compared to
$v_{\perp} (t-T_0)$.

\section*{CONCLUSION}

We considered the possibility of using long-term PAT observations
for the known millisecond pulsars in globular clusters to detect
intermediate-mass ($10^3-10^4 M_{\odot}$) black holes presumably
located at their centers.

–- The maximum signal delays over 5 years of observations were
estimated for several pulsars closest to the centers of the
corresponding globular clusters.

–- The detection of such a time delay by currently available
instruments for the pulsars known to date is not possible for a
black hole mass of $10^3 M_{\odot}$ and very problematic for a black
hole mass of $10^4 M_{\odot}$.

–- The pulse delay will have a negligible effect on the pulse
periods and their first derivatives compared to the current accuracy
of their measurements.

Nevertheless, note that using precise millisecond pulsar timing
methods in future (with an improvement in the resolution and
sensitivity of instruments, the detection of pulsars at angular
distances of a few fractions of an arcsecond from the globular
cluster centers, the detection of Galactic pulsars behind globular
clusters, a proper analysis of the low frequency noise, etc.) may
turn out to be one of the few tools for direct detection of IMBHs at
the centers of globular clusters.

\section*{ACKNOWLEDGMENTS}

This work was supported by the Russian Foundation for Basic Research
(project nos. 07-02-01051, 08-08-13734 and 07-02-00886), programs of
the Russian President (NSh-5579.2008.2) and the Presidium of the
Russian Academy of Sciences ''Origin, Structure and Evolution of
Universe Objects''). We are grateful to the International Space
Science Institute (ISSI, Bern, Switzerland) during the visits to
which we performed much of this work. We also wish to thank the
referees for a careful reading of the paper and helpful remarks. We
wish to thank A.Serber for the help in translating this paper in
English.

\pagebreak

\section*{REFERENCES}

1. G. C. Bisnovatyi-Kogan, B. I. Kolosov, and R. S. Churaev, Sov.
Astron. Lett. 6, 82 (1980)

2. R. van den Bosch, T. de Zeeuw, K. Gebhardt, et al., Astrophys. J.
641, 852 (2006)

3. G. Clark, T.Markett, and F. Li, Astrophys. J. 199, L93 (1975)

4. O. V. Doroshenko and S. M. Kopeikin, Mon. Not. R. Astron. Soc.
274, 1029 (1995)

5. A. Einshtein, Collected Works, in 4 vols. (Mir, Moscow, 1965) [in
Russian]

6. J. Frank and M. Rees, Mon.Not. R.Astron. Soc. 176, 633 (1976)

7. K. Gebhardt, R. M. Rich, and L. C. Ho, Astrophys. 634, 1093
(2005)

8. S. M. Kopeikin, Mon. Not. R. Astron. Soc. 305, 563 (1999)

9. S. M. Kopeikin and G. Schafer, Phys. Rev. D 60, 124002 (1999)

10. L. M. Krauss and T. A. Small, Astrophys. J. 378, 22 (1991)

11. T. I. Larchenkova and O. V. Doroshenko, Astron. Astrophys. 297,
607 (1995)

12. T. I. Larchenkova and S. M. Kopeikin, Astron. Lett. 32, 18
(2006)

13. T. I. Larchenkova and A. A. Lutovinov, Astron. Lett. 33, 455
(2007)

14. T. Maccarone, A. Kundu, S. Zepf, and K. Rhode, Nature 445, 183
(2007)

15. D. McLaughlin, J. Anderson, G. Meylan, et al., Astrophys. J.
166, 249 (2006)

16. B. J. McNamara, T. E. Harrison, and J. Anderson, Astrophys. J.
595, 187 (2003)

17. E. Noyola, K. Gebhardt, and M. Bergmann, ASP Conf. Ser. 352, 269
(2006)

18. S. Refsdal, Mon. Not. R. Astron. Soc. 128, 295 (1964)

19. M. V. Sazhin, in Proc. of the 11th Intern. Conf. on General
Relativity and Gravitation (Stockholm, Sweden, 1986), v. II, p. 519.

20. I. I. Shapiro, Phys. Rev. Lett. 13, 789 (1964)

21. E. Siegel, http://arxiv.org/abs/0801.3458 (2008)  

22. E. Turner, J. Ostriker, and J. Gott, Astrophys. J. 284, 1 (1984)

23. M. Vietri and J. Ostriker, Astrophys. J. 267, 488 (1983)

24. N. Wex, J. Gil, and M. Sendyk, Astron. Astrophys. 311, 746
(1996)

25. A. F. Zakharov, Gravitational Lenses and Microlenses (Yanus-K,
Moscow, 1997) [in Russian].


\pagebreak

\begin{table}[h!tbp]
\begin{center}
\renewcommand{\arraystretch}{1.1}
\caption{Spatial characteristics of the pulsars at minimum distances
from globular cluster (GC) center and signal time
delays}\label{tabl_clst}
\begin{tabular}{l|c|c|r|r|r|r}
\hline\hline

PSR (GC)  &  Distance  & $r_h$, pc & offset, & $d_m$, & $\Delta\tau_{3}$,& $\Delta\tau_{4}$,\\
          &  to GC, kpc&           & arcmin  & AU     & ns & ns \\
\hline
J0024-7204O (47 Tuc) & 4.1  & 3.33 &    0.06 & 14762  &  45  & 450 \\
J1748-2446C (Ter 5)  & 10.3 & 2.49 &    0.17 & 105070 &  0.9 & 8.9 \\
B1745-20 (NGC6440)   &  8.4 & 1.42 &    0.04 & 20162  &   24 & 240 \\
J1750-3703D (NGC6441)& 11.7 & 2.18 &    0.05 & 35103  &  8.0 & 80  \\
B1820-30A (NGC6624)  &  7.9 & 1.88 &    0.05 & 23702  & 17.5 & 175 \\
J1910-5959B (NGC6752)&  4.0 & 2.72 &    0.10 & 24002  & 17.1 & 171 \\
B2127+11D (M15)      & 10.2 & 3.18 &   0.019 & 11743  &   71 & 710 \\
\hline

\end{tabular}

\end{center}
\end{table}

\pagebreak

\begin{center}
\begin{figure}
\includegraphics[width=12cm]{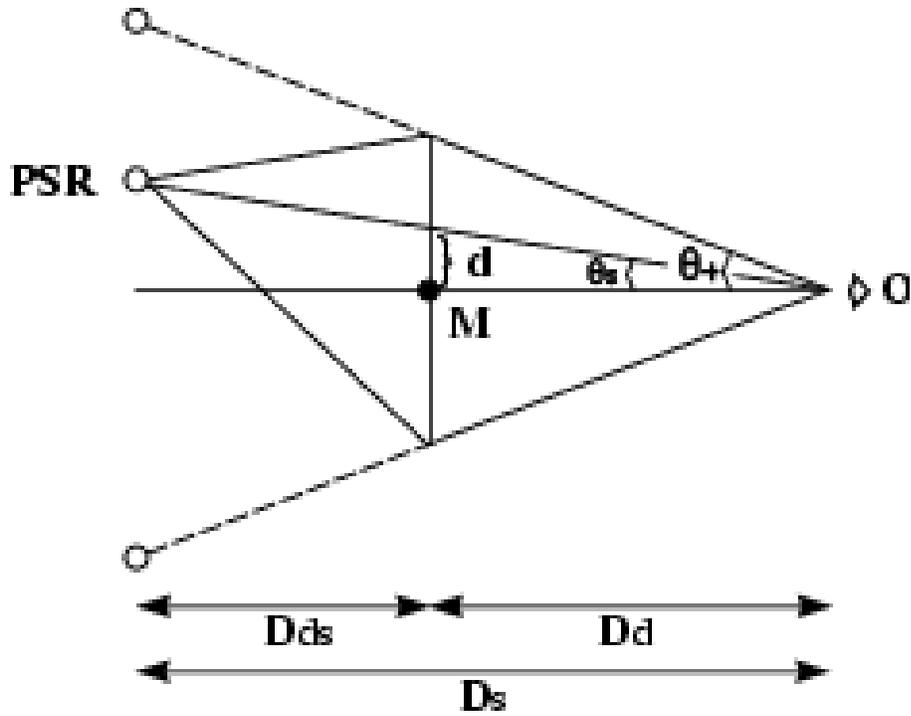}
\caption{Geometry of the problem under consideration: O is the
observer, PSR is the pulsar, and M is the black hole. For the
remaining notation, see the text.}
\end{figure}\label{geometry}
\end{center}

\pagebreak

\begin{center}
\begin{figure}
\includegraphics[width=12cm,bb=35 180 560 680]{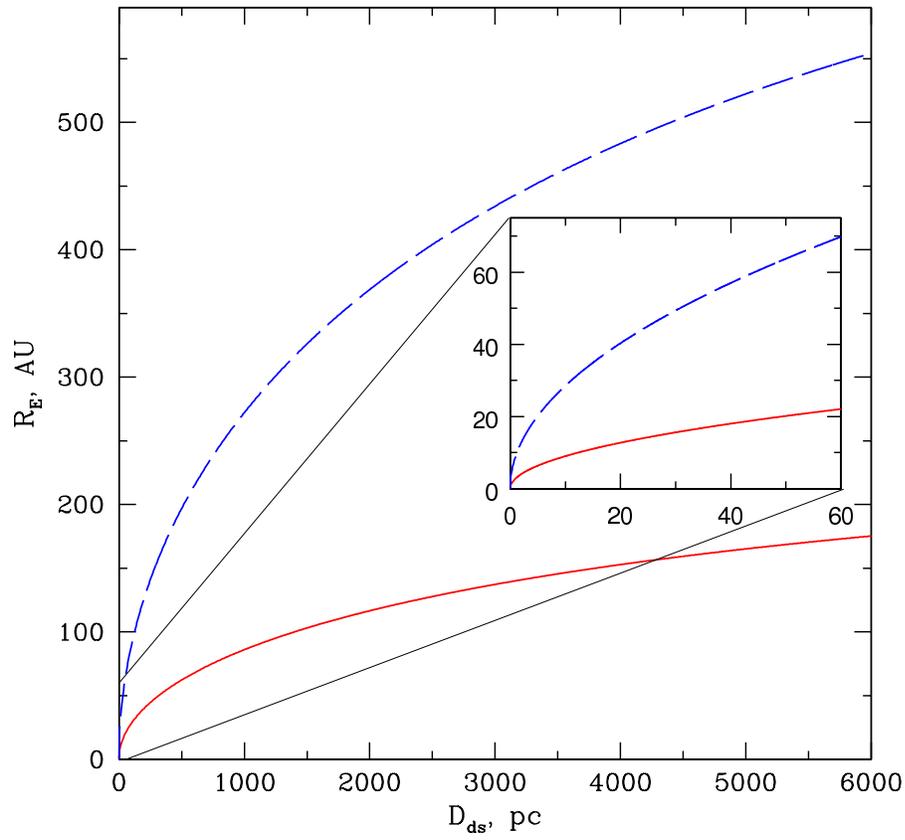}
\caption{$R_E$ versus distance between the black hole and the
pulsar: the black hole mass is $10^{3} M_{\odot}$ (solid line) and
$10^{4} M_{\odot}$ (dashed line).}
\end{figure}\label{re_dls}
\end{center}

\pagebreak

\begin{center}
\begin{figure}
\includegraphics[width=12cm,bb=35 180 560 680]{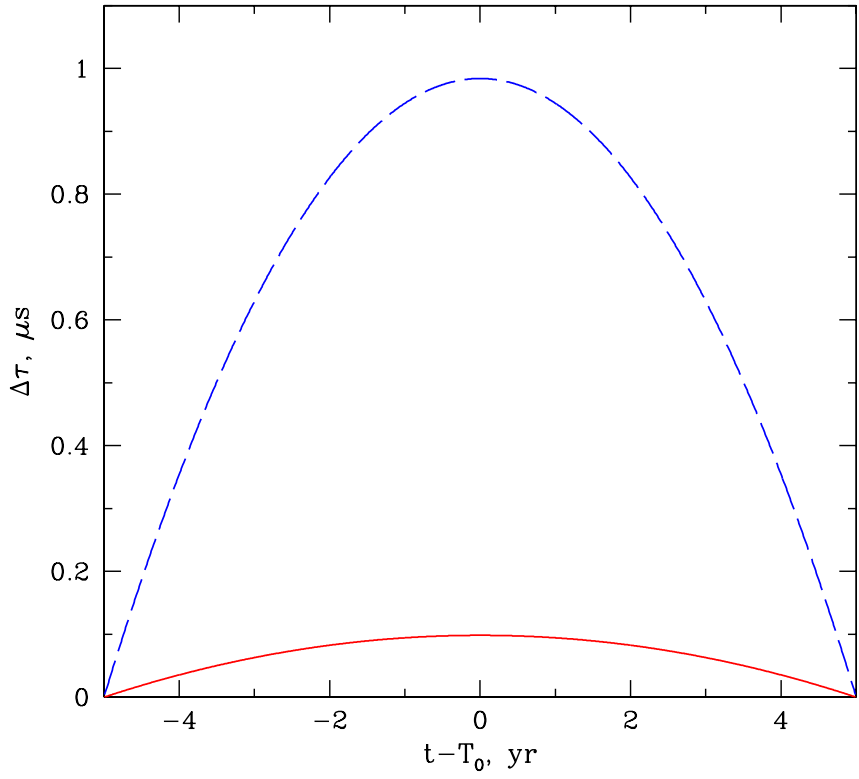}
\caption{Relative signal delay $\Delta \tau$ versus time $t–T_0$ (in
years) for two central black hole masses, $M_{BH} = 10^{3}
M_{\odot}$ (solid line) and $10^{4} M_{\odot}$ (dashed line).}
\end{figure}\label{deltaT}
\end{center}

\end{document}